%% file: manuscript_MGA.tex
\documentclass[preprint,12pt]{elsarticle}

\usepackage{amssymb}
\usepackage{amsmath}

\usepackage{lineno}

\usepackage{standalone}
\usepackage{multicol}
\usepackage{multirow}
\usepackage{siunitx}
\usepackage{soul}
\usepackage{subcaption}

\usepackage{tabularx}
\usepackage[table]{xcolor}
\usepackage{booktabs}
\usepackage{float}
\usepackage[section]{placeins}
\usepackage[hidelinks]{hyperref}
\usepackage{tikz}
\usetikzlibrary{positioning}
\usepackage[utf8]{inputenc}
\usepackage [style = english]{csquotes}
\usepackage{graphicx}                 
\graphicspath{{figures/}}     
\MakeOuterQuote{"}
\usepackage{layouts}
\biboptions{sort&compress}
\journal{}
\newcommand{\co}{CO\textsubscript{2}}
\newcommand{\Mco}{MtCO\textsubscript{2}/a}

\graphicspath{ {figures} }
\begin{document}

\begin{frontmatter}



\title{Near-optimal solutions for carbon capture, conversion, storage, and removal strategies}


\author[a,2]{Sina Kalweit} 
\ead{sk@mpe.au.dk}
\author[a,2]{Ricardo Fernandes}
\author[a]{Alberto Alamia}
\author[a,2,5]{Marta Victoria}

\affiliation[a]{organization={Department of Mechanical and Production Engineering, Aarhus University},
            addressline={Katrinebjergvej 89}, 
            city={Aarhus},
            postcode={8200}, 
            country={Denmark}}

\affiliation[2]{organization={Novo Nordisk Foundation CO2 Research Center},
            addressline={ Gustav Wieds Vej 10}, 
            city={Aarhus},
            postcode={8000}, 
            country={Denmark}}

%
\affiliation[5]{organization={Department of Wind and Energy Systems, Technical University of Denmark},
            addressline={Elektrovej, 325}, 
            city={Lyngby},
           postcode={2800}, 
            country={Denmark}}
\begin{abstract}
Achieving climate neutrality in Europe requires rapid electrification alongside carbon management strategies for residual emissions. Existing analyses of the European energy system often focus on collocated carbon capture and geological sequestration, with limited attention to the interactions among carbon capture and utilization, transport, sequestration, and diverse carbon dioxide removal (CDR) options. Moreover, existing literature focuses on discussing the optimal, neglecting that near-optimal solutions might provide very different system configurations at a marginal higher cost. Here, we integrate afforestation, biochar, enhanced rock weathering, and perennialization into a sector-coupled European energy system model (PyPSA-Eur) clustered to 39 nodes with 750 aggregated time steps. We explore their contributions using a Modelling to Generate Alternatives (MGA) approach. The approach combines minimization, maximization, and random vectors to explore the near-optimal solution space for up to 5\% increased total system costs. Our results show that, in a carbon-neutral system, multiple configurations of carbon management options can achieve net-zero emissions with only marginal cost increases. We find that a 5\% total system cost increase is sufficient to accommodate the full spectrum from zero to full deployment of the individual CDR options, as well as a wide range of synthetic fuel use across different fuel types. Increased reliance on CDR options offers no clear cost advantage compared to greater utilization of synthetic fuels.
\end{abstract}




\end{frontmatter}


\section{Introduction}
The European Union aims for climate neutrality by 2050, and cost-effective decarbonization requires a large increase in renewable electricity from wind and solar to replace fossil fuels with direct and indirect electrification across the energy system \cite{Victoria2022,Luderer2021}. On top of that, the IPCC's 6th Assessment Report and the World Energy Outlook of the IEA state that carbon dioxide removal (CDR) is necessary to counterbalance hard-to-abate emissions and consequently to reach net-zero greenhouse gas emissions \cite{IPCC2023, IEA2022,Rodriguez2017}. Previous studies have shown that, in cost-optimal solutions, carbon capture and sequestration technologies are used to reach this goal \cite{Victoria2022, Tatarewicz2021,Creutzig2019,VanVuuren2018,Markkanen2024,Fuhrman2023,Strefler2021, IPCC5}. Bioenergy with carbon capture and storage (BECCS) has been widely used in many analyses \cite{IPCC5} despite the concerns regarding its sustainability \cite{Creutzig2019,VanVuuren2018}, and direct air capture (DAC) combined with underground sequestration has also been included in many sector-coupled energy system models \cite{Victoria2022,Fasihi2019}. Alternative CDR strategies, such as afforestation, enhanced rock weathering (ERW), conversion of seasonal to perennial crops (perennialization), or biochar, have been represented in global integrated assessment models (IAMs) \cite{Fuhrman2023, Strefler2021} or European models \cite{Markkanen2024}. Recently, we extended the open sector-coupled energy system model of Europe, PyPSA-Eur, that cost-optimizes capacity and dispatch, to include a representation of these alternative CDR strategies with high spatial resolution \cite{Fernandes}.

Previous results are based on minimizing system costs. However, additional considerations for socio-economic, political, or environmental reasons may make a slightly costlier solution more attractive \cite{Lombardi2025}. Some studies address this by exploring the near-optimal solution space, where a slight increase in costs offers a wide variety of possible solutions and allows understanding how robustly a technology is chosen \cite{Neumann2021,Millinger2025,Pedersen2021, Pickering2022,Greevenbroek2025}. Different modelling to generate alternatives (MGA) algorithms have been proposed to explore that space \cite{Lau2024}. So far, the focus has mostly been on electricity generation and conversion technologies such as wind, solar, batteries, or hydrogen, showing the flexibility to vary capacities between renewable options \cite{Pedersen2021,Pickering2022,Greevenbroek2025} and the possibility to find solutions that maximize other objectives such as uniform distribution of generation capacities among regions \cite{Neumann2021}. Recently, Millinger et al. \cite{Millinger2025} explored near-optimal solutions to investigate the role of biomass in net-zero systems and found that a system cost increase of 14\% is necessary to completely replace BECC with DAC. However, an overview of the near-optimal solution space with a focus on diverse carbon capture, conversion, sequestration, and removal strategies is still missing. It remains unclear how variable the use of those strategies is, especially whether some of them prove to be necessary for the transition or if their use is optional. 

This study focuses on evaluating the possible solution space for using different carbon capture, conversion, sequestration, and removal strategies: DAC, point-source carbon capture (from process emissions, fossil or biomass combustion), \co-to-X conversion pathways (liquid hydrocarbons, methanol, methane), underground sequestration, ERW, afforestation, biochar, and perennialization for a net-zero European energy system. We use an MGA method combining minimization, maximization, and random vectors \cite{Lau2024} applied to the PyPSA-Eur model with extended CDR alternatives \cite{Fernandes}. Compared to the cost-optimal solution that maximizes afforestation, ERW, and perennialization \cite{Fernandes}, we find that even a 1\% total system cost increase allows a wide range of carbon management tools, and that perennialization and ERW can be excluded. However, afforestation, point-source carbon capture from fossil fuels, and \co\ underground sequestration cannot be completely replaced even with a 5\% system cost increase. Furthermore, a 1\% cost increase is enough to introduce biochar and synthetic methane production, which were not part of the optimal solution, and to allow a wide range of synthetic oil usage. 

\section{Methods}
\subsection{Energy system model}
This study uses PyPSA-Eur, an open European energy system model of the electricity, heating, industry, and transport sectors \cite{Neumann2023} to create a cost-optimal solution for a net-zero scenario in 2050. Regions are clustered to 39 nodes, and the available hourly time resolution is clustered to 750 time steps based on a segmentation approach, which was shown to be more accurate than a corresponding uniform aggregation \cite{Greevenbroek2025}.
We followed a greenfield approach with perfect foresight for the full year. We decided to assess the impact of removal, capture, conversion, and sequestration independently of wide infrastructure changes such as the expansion of transmission lines or building a H\textsubscript{2}-network and \co\ network. 
The following demands are set exogenously: oil for aviation, agricultural machinery oil, naphtha as feedstock for industry, methanol for shipping demand, and industrial heat demand covered by biomass and gas. The amount of process emissions is also set exogenously.
\subsection{Carbon management}
In our model, carbon management measures contributing to overall carbon neutrality can be grouped into four categories: capture, conversion, sequestration, and CDR strategies, see Tab. \ref{tab:CDR}. Carbon dioxide can be captured from air via direct air capture (DAC) or from processes that emit or release \co\ as a side product, such as: combined heat and power (CHP) plants powered by gas or solid biomass, producing hydrogen from gas via steam methane reforming (SMR), using solid biomass for medium-temperature and gas for high-temperature heat demand in industry, \co\ release during chemical reactions in industry (process emissions), upgrading biogas to pure biomethane, and converting biomass to methanol or oil. Throughout the study, ‘oil’ refers to a generic liquid hydrocarbon energy carrier that may originate from synthetic \co-based pathways, biomass-derived processes, or fossil resources, and may be used either as a fuel for aviation and agricultural machinery oil or as a refinery feedstock for industry. For every process involving carbon capture (CC), it is assumed that 90\% of the released \co\ can be captured. Processes based on fossil fuels or industrial process emissions lead to net \co\ emissions because 10\% of their emissions remain uncaptured. In contrast, biogas- and biomass-based processes are assumed to be carbon-neutral, as the released \co\ was previously absorbed from the atmosphere during biomass growth. Consequently, the amount of captured \co\ usable for conversion or sequestration is not equivalent to its contribution to reducing atmospheric \co\ or the overall emission balance, see \ref{Sup:carbonmanagement}.

The captured \co\ can be converted into carbonaceous fuels, commonly referred to as carbon capture and utilization (CCU), or sequestered underground for long-term removal from the atmosphere. There are three conversion possibilities in PyPSA-Eur using \co\ hydrogenation: to methane via the Sabatier reaction, to liquid hydrocarbons, which can be used as an oil substitute, via the Fischer-Tropsch process, or to methanol. Another option for the captured \co\ is sequestration, which is a long-term storage with the goal of preventing its reentry into the atmosphere. Only the combination of \co\ sequestration with \co\ capture from the atmosphere (DAC) or biogenic \co\ capture counts as a true net-negative contribution, which can offset emissions from industrial processes and fossil fuels. This effectively means that the maximum potential assumed for \co\ sequestration limits the negative emissions that can be attained with this strategy.

Moreover, we have recently modeled additional \co\ removal strategies that do not depend on underground \co\ sequestration \cite{Fernandes}. These carbon-dioxide removal (CDR) strategies are afforestation, where \co\ from the atmosphere is both captured and stored in trees, conversion of seasonal to perennial crops and green biorefining, where deep root systems and the less disturbed soil take up more \co and additional biogas is produced, biochar, the carbon-rich and stable remains of pyrolysis of biomass, and enhanced rock weathering (ERW), the increase of \co\ uptake by silicate-rich rocks by grinding them to powder and spreading them over a large surface area.

\setlength{\extrarowheight}{3pt}

\begin{table}[h]
  \small
    \centering
    \begin{tabularx}{\linewidth}{l}\toprule
        \textbf{Capture} \\\midrule
        Direct air capture  \\
        \textit{point-sources:} \\
        \hspace{3mm}Combined heat and power plant (CHP) using gas or solid biomass \\
        \hspace{3mm}Steam methane reforming (SMR) \\
        \hspace{3mm}Solid biomass for medium-temperature heat demand in industry  (<500  \textdegree C) \\
        \hspace{3mm}Gas for high-temperature heat demand in industry (>500  \textdegree C) \\
        \hspace{3mm}Industrial emissions due to chemical reactions \\
	      \hspace{3mm}Biogas to gas upgrading \\
        \hspace{3mm}Biomass to liquid \\
      	\hspace{3mm}Biomass to methanol  \\\midrule
        \textbf{Conversion}  \\\midrule
        Sabatier: H\textsubscript{2} and \co\ to methane \\
        Fischer-Tropsch: H\textsubscript{2} and \co\ to liquid hydrocarbons \\
        Methanol synthesis: H\textsubscript{2} and \co\ to methanol  \\\midrule
        \textbf{Sequestration} \\\midrule
        \co\ sequestered underground \\\midrule
        \textbf{Removal (CDR)} \\\midrule
        Conversion of seasonal to perennial crops \\
        Afforestation \\
        Enhanced rock weathering \\
        Biochar \\
\bottomrule
    \end{tabularx}
    \caption{Overview of carbon management strategies in PyPSA-Eur. The capacities of these technologies and strategies are selected as decision variables to explore the near-optimal solution space.
    }
    \label{tab:CDR}
\end{table}

In the model, the capacities of afforestation, biochar, ERW, and perennialization are node-wise restricted due to land use and climate restraints. \co\ underground sequestration is also spatially resolved based on the availability of offshore salt deep salt caverns and depleted hydrocarbon reservoirs. A limit of 200 \Mco\ is imposed to avoid overreliance on underground sequestration, in contrast to decarbonizing sectors as discussed \cite{Hofmann2025}. DAC and point-source capture are only indirectly limited by the \co\ sequestration potential, the biomass potential, and industrial demands.\\

\subsection{MGA method}
MGA algorithms explore the near-optimal feasibility space of an optimal solution by introducing a relaxed cost constraint and an alternative objective function \cite{DeCarolis2011}.
The objective function varies selected decision variables, typically the capacities of technologies such as wind power or CHP plants, so that successive solutions differ significantly, enabling efficient exploration of the solution space. 

Previous studies have used minimization and maximization (Min/Max) \cite{Neumann2021, Millinger2025}, Hop-Skip-Jump (HSJ) \cite{DeCarolis2011}, combinations of HSJ and Min/Max \cite{Pickering2022}, or Modelling All Alternatives (MAA) \cite{Pedersen2021}, with MAA typically limited to ten decision variables \cite{Pedersen2021}. Lau et al. \cite{Lau2024} note limitations of Min/Max and HSJ and propose combining Min/Max with Random vectors for broader, faster exploration. Based on this, we apply a combination of Min/Max and Random vectors approaches using the spatially aggregated capacities of carbon management options as decision variables, see Tab. \ref{tab:CDR}.

The cost-optimal solution $z^{*}$ is found by:
\begin{align}
&\min z = f(x) \nonumber \\
s.t. \nonumber \\
&g_i(x) \leq 0,\, i=1,\ldots,m\\
&h_j(x) = 0,\, j=1,\ldots,p \nonumber
\end{align}
where $z=f(x)$ is the total system cost and $g_i(x), h_j(x)$ are physical constraints. A detailed description of those constraints is provided in Appendix S15 of \cite{Victoria2022}. 

A relaxed cost constraint is imposed for each MGA solution $k$:
\begin{equation}
f(x^{(k)}) \leq z^{*}\cdot(1+\epsilon)
\end{equation}
where $\epsilon$ specifies the allowable relative increase in system cost. Decision variables $x^{(k)}$ are weighted by a vector $v$, sampled from a uniform distribution over [-1,1] or the integer set [-1,0,1], giving the MGA problem:
\begin{align}
&\min/\max v \cdot x^{(k)} \nonumber \\
s.t. \nonumber \\
&g_i(x) \leq 0,\, i=1,\ldots,m\\
&h_j(x) = 0,\, j=1,\ldots,p \nonumber \\
&f(x^{(k)}) \leq z^{*}\cdot(1+\epsilon) \nonumber
\end{align}
Following Lau et al. \cite{Lau2024}, about one quarter of $v$ is sampled from the uniform distribution. Convergence is assessed via volume expansion, as described in Supplementary Note 1.2.

\section{Results}
\subsection{Carbon management across the near-optimal solution space}
\begin{figure}[h]
\includegraphics[width=\textwidth]{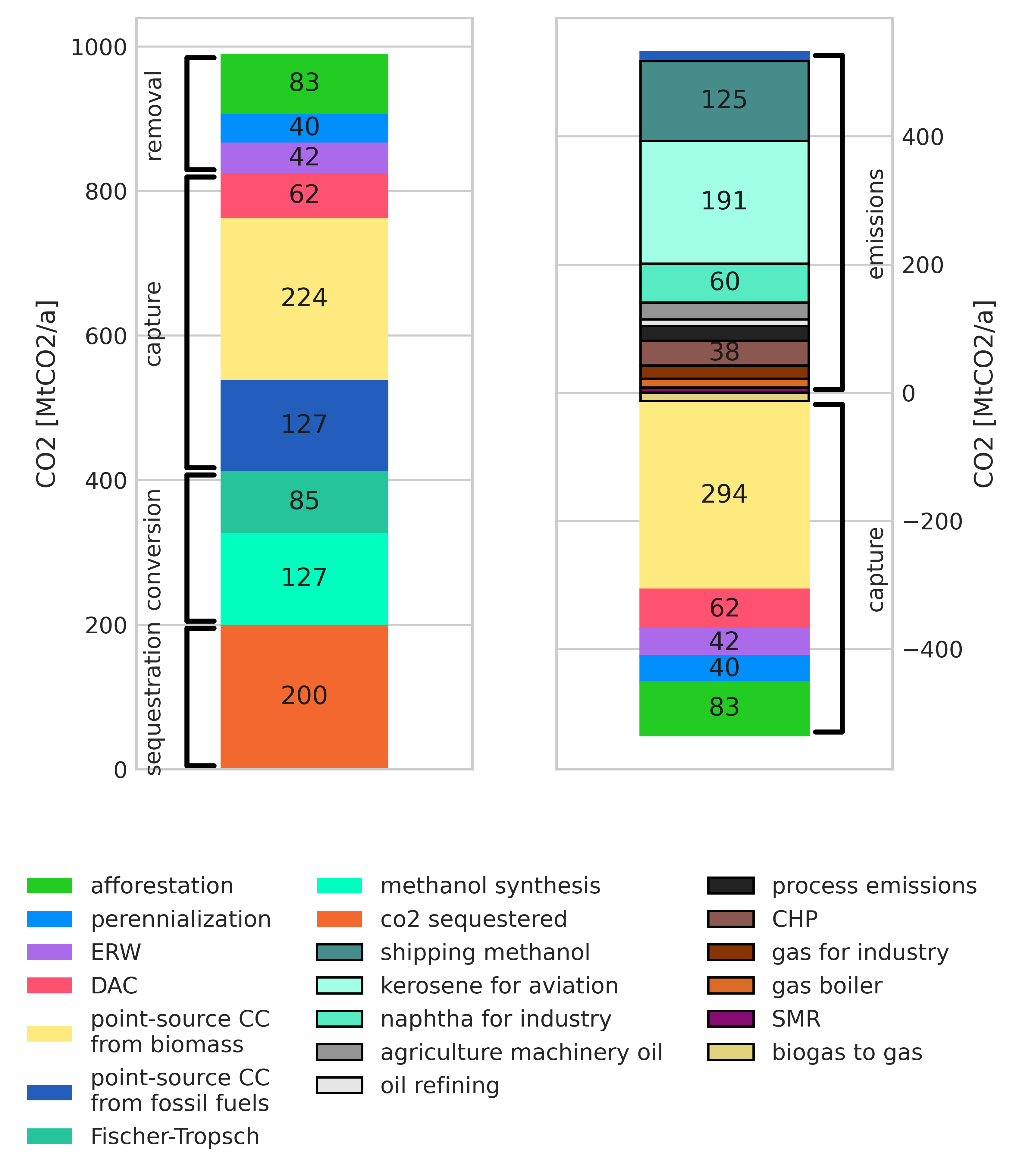}
\caption[]{Optimal solution on which near-optimal solutions are based. Overview of the \co\ that is captured from point-source emissions or conversion processes and is available for underground sequestration or conversion to synthetic fuels and \co\ removal (left) and how carbon capture balances \co\ emissions in the atmosphere (right). }
\label{Fig:opt}
\end{figure}

Although multiple carbon management options exist, the optimal solution only uses a subset of them, as shown in Fig. \ref{Fig:opt}. Among the CDR strategies, it fully exploits the potential of afforestation, perennialization, and ERW, while excluding biochar. Biochar is excluded because the available biomass is preferentially used to produce bio-oil and supply heat during peak demands \cite{Fernandes}. Among capture technologies, 350 \Mco\ are captured by point-sources; in contrast to 62 \Mco\ by DAC. Of this captured \co, 127 \Mco\ are converted to synthetic methanol, 85 \Mco\ to synthetic oil, and none to synthetic methane. The remaining 200 Mt of captured \co\ are sequestered underground, exhausting the full underground sequestration potential.
Within the overall system balance, aviation is the largest emission source with 191 \Mco\ (Fig. \ref{Fig:opt}), and it strongly influences the system's carbon management. The exogenous demand of aviation and the chemical industry for oil must be supplied by fossil oil, biomass-to-liquids, or synthetic oil production via Fischer-Tropsch, but the cost of the fossil option is significantly lower than the other two. Moreover, our conservative assumption of the biomass potential limits the biomass-to-liquids route. In the optimal scenario, 85 \Mco\ of captured \co\ is used to produce synthetic oil. The remaining oil demand is supplied by fossil oil, whose emissions are offset by CDR.\\

\begin{figure}[h]
\includegraphics[width=\textwidth]{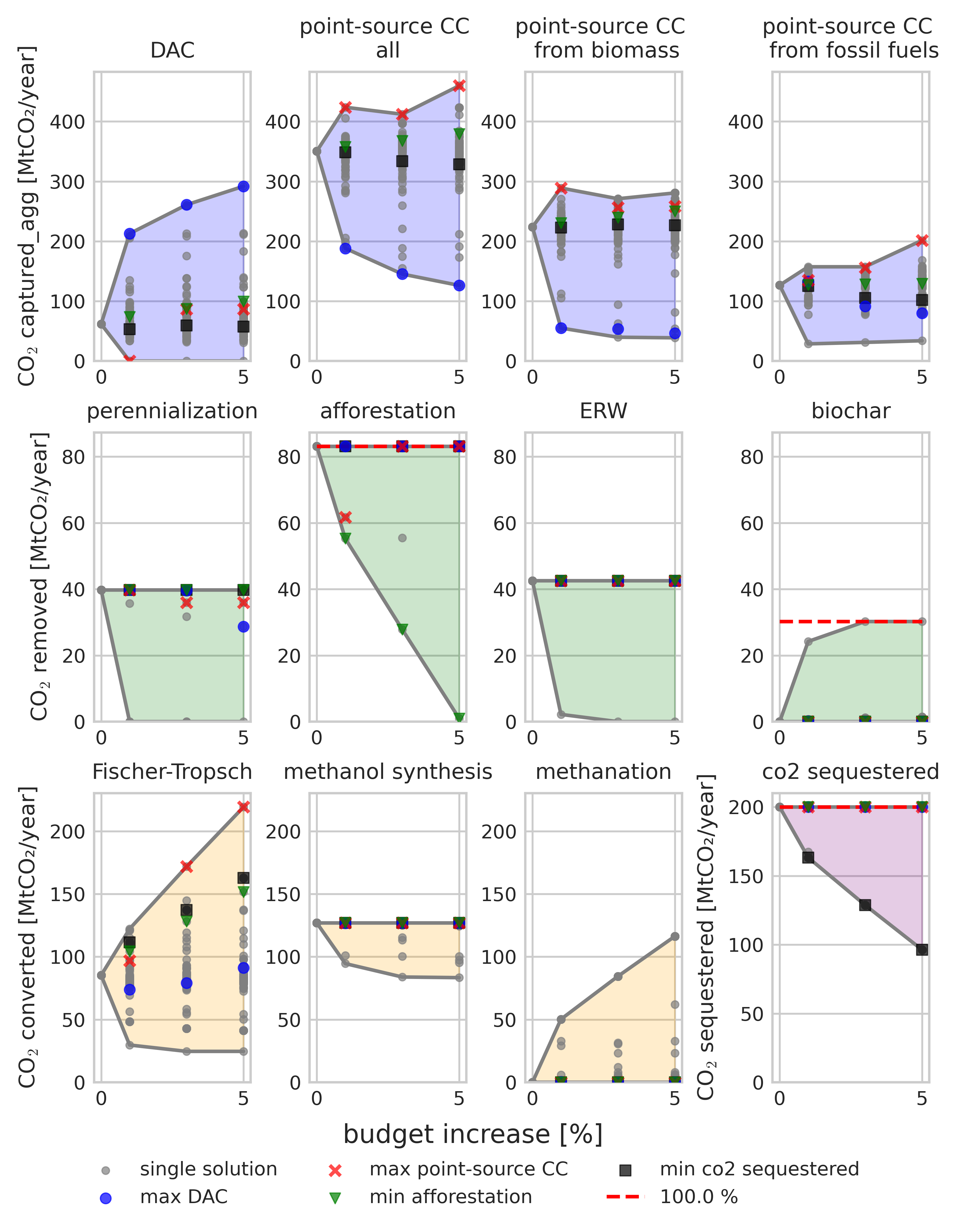}
\caption[]{Solution space for carbon management strategies (see Tab. \ref{tab:CDR}). Point-source CC is the sum of all capture technologies except for DAC. Solutions were obtained for a total system cost increase of 1\%, 3\%, and 5\% based on the optimal solution. }
\label{Fig:comp}
\end{figure}

The near-optimal solution space is quite flat, allowing significant variations on the selected strategies for carbon management, see Fig. \ref{Fig:comp}.
Across CDR strategies, a 1\% increase in total system costs allows utilization of up to 80\% (24 \Mco) of biochar's total potential, and solutions that do not use perennialization or reduce removal by ERW to 2 \Mco. In contrast, a 5\% budget increase is necessary to reduce afforestation to 1 \Mco, see Fig. \ref{Fig:comp}. Afforestation is very attractive for the system because it does not have any energy requirements and its marginal cost is lower than those of ERW, perennialization, or biochar, see Tab. \ref{tab:cdrs_parameters}. \\

\begin{figure}[h]
\centering
\includegraphics[width=\textwidth]{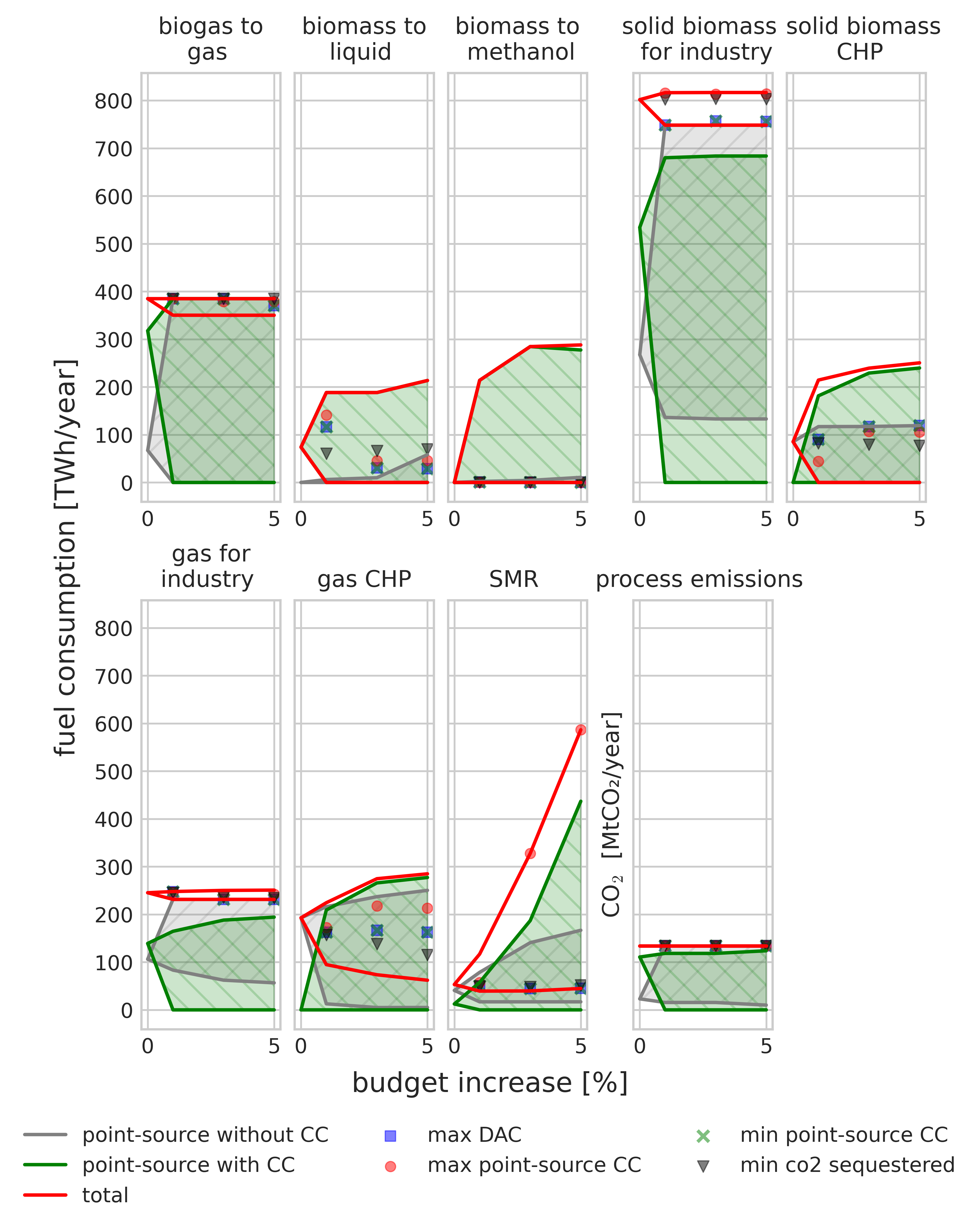}
\caption[Solution space: Point-source with and without carbon capture]{Solution space for point-source utilization with and without carbon capture. Solutions were obtained for a total system cost increase of 1\%, 3\%, and 5\% based on the optimal solution. While biomass for industry demand is set exogenously, using it with carbon capture lowers the efficiency of the process, resulting in a 1.1\% higher biomass demand. Single solutions are shown for technologies with and without carbon capture combined.}
\label{Fig:compPS}
\end{figure}

For capture technologies, individual fossil-based point-source capture technologies are mutually interchangeable within a 1\% budget increase, as are biomass-based point-source capture technologies. Only biomass-based point-source capture leads to net \co\ removal when combined with \co\ sequestration, whereas fossil-based point-source capture merely reduces emissions. Point-source carbon capture is always necessary, and ranges between 188 and 424 \Mco. For a 5\% budget increase, the lower boundary of point-source capture decreases to 126 \Mco. The exogenously-determined solid biomass for industry utilization shows an especially high capture potential compared to the other point-source technologies, see Fig. \ref{Fig:compPS}. DAC captures up to 213 \Mco\ for a 1\% increase and up to 292 \Mco\ for a 5\% increase. The lowest boundary of the combined amount of captured \co\ from point-source and DAC is 384 \Mco\ for a 5\% budget increase.  \\

Across conversion technologies, production of synthetic oil via Fischer-Tropsch shows the widest range, spanning from utilizing 30 \Mco\ to 122 \Mco\ for only a 1\% budget increase and up to 219 \Mco\ for a 5\% budget increase. The other two conversion technologies show lower ranges of variability. For a 5\% budget increase, synthetic methanol production can be reduced to utilizing 83 \Mco , and synthetic methane, which is not used in the optimal solution, can use up to 116 \Mco. \\

For underground sequestration, the lower-bound solution corresponds to 163 \Mco\ for a 1\% budget increase. Even with a 5\% budget increase, \co\ sequestration is only reduced by at most 48\% (96 \Mco), demonstrating how important this strategy is to attain carbon neutrality.

While the near-optimal solution space shows a wide range of possibilities, all near-optimal solutions must be carbon-neutral, which means that the sum of \co\ removals is equal to the sum of \co\ emissions. Together with requirements for individual technologies and strategies, this constrains the flexible use of carbon management options in individual solutions.
The following sections explore these interdependencies by examining different extreme scenarios where specific technologies or strategies are either maximized or minimized, and how these choices impact the utilization of other carbon management options. 

\FloatBarrier

\subsection{Synthetic oil can partially reduce the need for CO\textsubscript{2} underground sequestration}
\begin{figure}
\includegraphics[width=\textwidth]{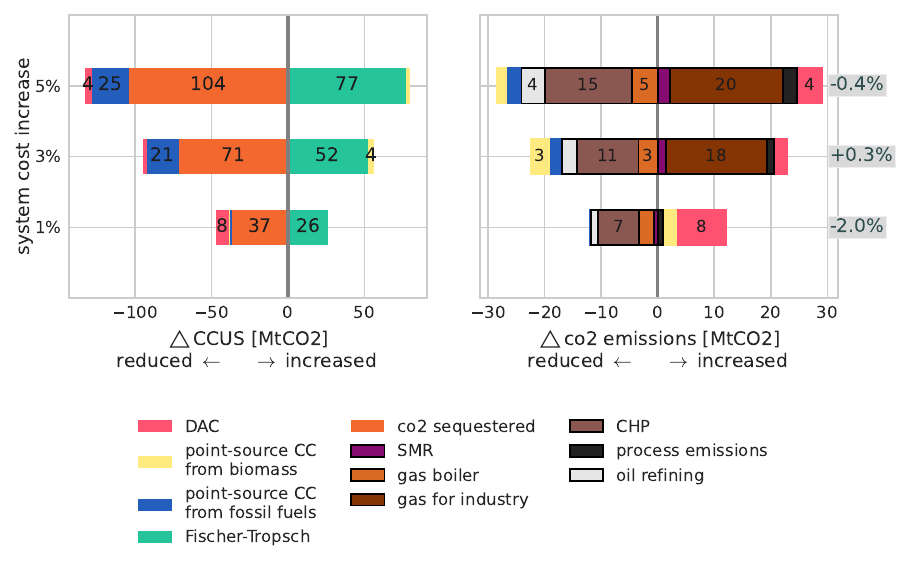}
\caption[Solution space: Reducing \co\ sequestration underground]{Differences of carbon management strategies utilization for solutions with minimal \co\ sequestration underground use compared to the optimal solution. A reduction of capture technologies shown on the left figure (e.g. DAC) translates into an increase in \co\ emissions shown on the right figure. The grey boxes for the \co\ balance show the total amount of emissions being increased or reduced compared to the optimal solution.}
\label{Fig:minseq}
\end{figure}
A minimum of 96 \Mco\ underground sequestration is required even at a 5\% total system cost increase, see Fig. \ref{Fig:minseq}. In these scenarios, the CDR strategies perennialization, ERW, and afforestation are fully utilized to maintain 165 \Mco\ of \co\ removal, while biochar deployment is limited by biomass availability.
Minimizing underground sequestration leads to a reduction in fossil fuel use by increasing synthetic oil production via Fischer-Tropsch synthesis. This requires an additional amount of 77 \Mco\ for a 5\% budget increase, which is lower than the corresponding reduction of underground sequestration. As a result, the need for DAC and point-source capture is reduced.\\

DAC and point-source capture from fossil fuels decrease, while point-source capture from biomass for industry slightly increases at higher budget levels, see Fig. \ref{Fig:compPS}. The deployment of gas-fired and biomass-fuelled CHP plants, as well as gas-based boilers, is partially reallocated to biomass-based boilers, using biomass that is otherwise utilized in biomass-to-liquid processes and CHP generation in the optimal solution. Increased Fischer-Tropsch production reduces oil refining and associated emissions but raises hydrogen demand. This additional demand is met through increased hydrogen electrolysis. Additional wind and solar capacity is installed to supply the electricity required for the increased electrolysis demand, see Fig. \ref{Fig:othersectors}.\\ 

Across these scenarios, natural gas use declines from 179 TWh in the optimal solution to 126 TWh, 92 TWh, 65 TWh for 1\%, 3\%, 5\% budget increase, respectively. At higher budget levels, less \co\ is captured from gas and more from biomass, the latter allowing net negative emissions.

\subsection{Minimizing CDRs increases DAC, biomass-based point-source capture, and Fischer-Tropsch}
In the optimal solution, the CDR strategies afforestation, ERW, and perennialization are used to their maximum potential. With \co\ sequestration underground also fully exploited, CDR strategies can only be replaced by the only other net \co\ removal method, biochar, or by reducing emissions by decreasing fossil fuel use. Across the near-optimal solutions, minimizing afforestation for the 5\% budget increase is the scenario with the lowest amount of CDRs used.  \\

\begin{figure}
  \includegraphics[width=\textwidth]{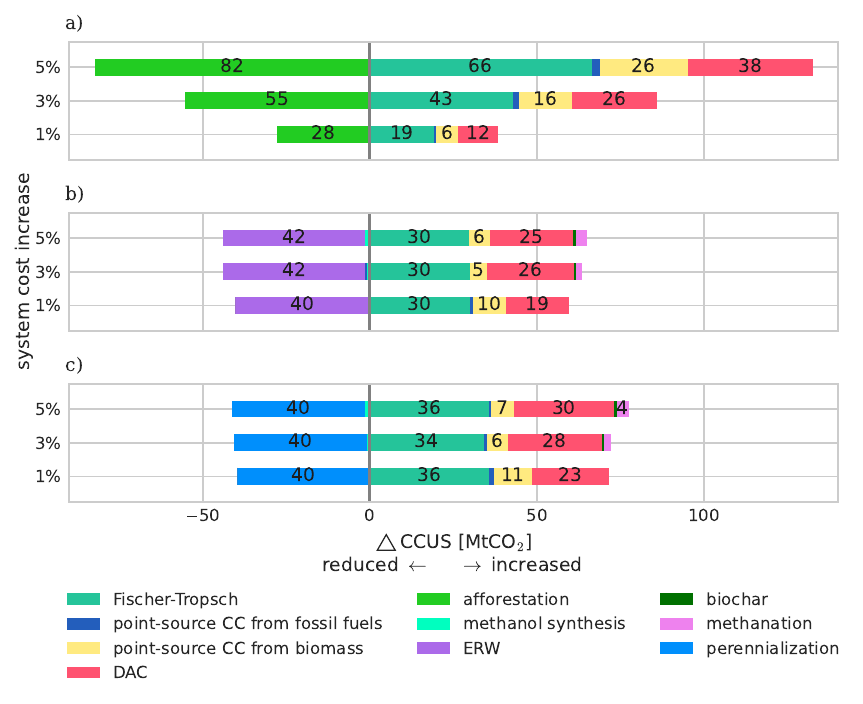}
\caption[Solution space: reducing CDRs]{Solutions for minimizing a) afforestation, b) ERW, c) perennialization. Values indicate changes in capturing, removal, sequestration, and conversion compared to the optimal solution.}
\label{Fig:ex1}
\end{figure}

Minimizing afforestation, ERW, or perennialization leads to a decrease of total emissions by up to 18 \Mco\ (3.3\%) compared to the optimal solution, see Fig. \ref{Fig:ex1}. The remaining emissions that were previously offset by CDR strategies are now balanced by an increased use of carbon capture and conversion to synthetic oil via Fischer-Tropsch, see Fig. \ref{Fig:ex1}. For a 5\% budget increase, the amount of carbon capture by DAC rises by up to 38 \Mco\ and biomass-based point-source rises by up to 26 \Mco\ additionally captured. 

The limited substitution of CDR strategies by biochar is primarily due to biomass constraints. At least 72\% of the biomass potential is exogenously appointed for high-temperature heat demand in industry (see solid biomass to industry in Fig. \ref{Fig:compPS}). Assuming that the remaining biomass potential were completely used for the production of biochar, this corresponds to a \co\ removal potential of 38 \Mco. While this is higher than the biochar potential based on land availability of 30 \Mco (see Tab.S2), biochar competes with biomass used to produce oil or methanol and as a fuel for boilers and CHP plants. As a result, the system relies more heavily on additional synthetic fuel production to reduce fossil oil consumption and balance emissions.\\

\FloatBarrier
\subsection{Influence on other sectors}
\begin{figure}[h]
\centering
\includegraphics[width=\textwidth]{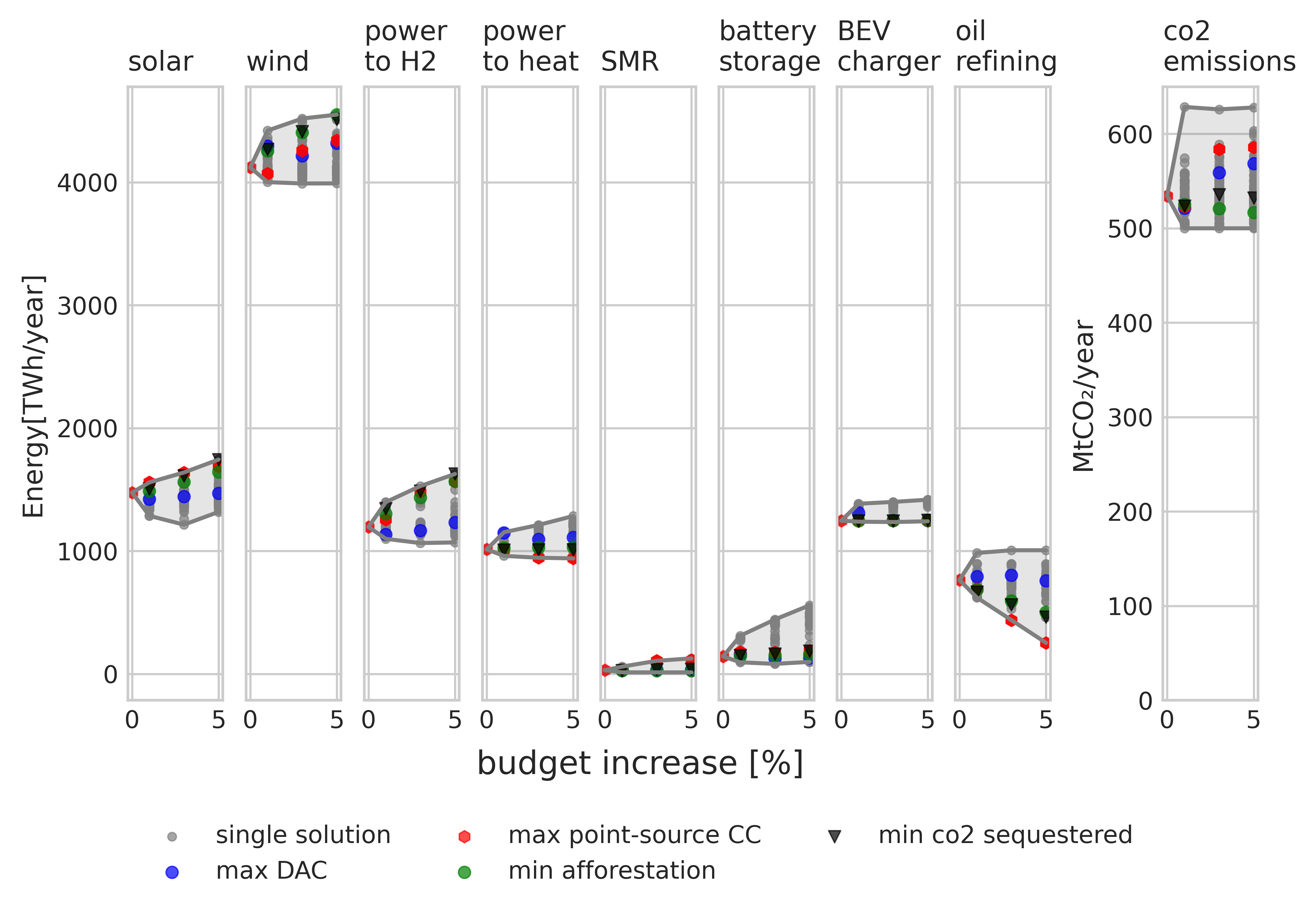}
\caption[Solution space for other sectors]{Solution space for other sectors. Solutions were obtained for a total system cost increase of 1\%, 3\% ,and 5\% compared to the cost-optimal solution.}
\label{Fig:othersectors}
\end{figure}
Technologies that were not used as decision variables for the near-optimal solution space still vary across solutions, see Fig. \ref{Fig:othersectors}. As discussed before, oil refining is directly influenced by the amount of synthetic oil or bio-oil in the system. Synthetic fuel production increases hydrogen and electricity demand. Different scenarios impact the total emissions in the system; in general, high DAC use increases emissions, see Fig. \ref{Fig:ex3} and minimization of \co\ underground sequestration or CDR strategies reduces emissions, see Fig. \ref{Fig:minseq} and \ref{fig:exewper}. In the optimal solution, 534 \Mco\ were emitted. In the near-optimal solutions, this ranges from 500 to 629 \Mco\ for a 1\% increase.
\section{Discussion}
This study found that, while the optimal solution of a net-zero European energy system fully utilizes several carbon management strategies (perennialization, ERW, afforestation, and underground sequestration) and excludes others (biochar and synthetic methane), an increase in total system costs by only 1\% allows a wide range of carbon management tools. Although the near-optimal solution space shows a preference for the full use of available underground sequestration and CDR strategies, a few solutions feature substantial reductions in CDRs, and previously unused carbon management tools can be introduced. Some carbon management strategies are difficult to replace, namely underground sequestration, carbon capture from point-sources based on fossil fuels or process emissions, and afforestation. Furthermore, near-optimal solutions change in their amount of total emissions, with the possibility of reducing emissions further. \\

In this paragraph, we compare our results with those in the relevant literature. Our optimal solution includes a low usage of DAC, which is in agreement with the negligible contribution from this technology found by Millinger et al. \cite{Millinger2025} in a European system and Fuhrman et al. \cite{Fuhrman2023} for Europe in a global system. In contrast, Markkanen et al. \cite{Markkanen2024} report DAC use combined with underground sequestration ranging from over 250 \Mco\ to more than 800 \Mco\ \cite{Markkanen2024}, which is much higher than the 308 \Mco\ we see in the near-optimal solution space with a 5\% slack. Their high usage of DAC can be explained by their lower cost assumption (34\% lower than in our studies) and their higher assumption for underground sequestration potential (78-175 Gt\co\ cumulative for 2025-2065). Markkanen et al. investigate six different scenarios which, on average, use a higher CDR potential than we assume here of 1.1 Gt\co/a. They include bioenergy with carbon capture, biochar, ERW, underground sequestration, afforestation and reforestation, and solid-carbon sequestration, and their scenarios assume at least ten times the amount of underground sequestration potential, around 272 \Mco\ removal from reforestation, and up to 176 \Mco\ removal from soil carbon sequestration. Biochar and ERW potential in our model are limited by conservative assumptions about crop land use and performance. However, they qualitatively agree with our optimal solution in that their scenarios up to 2050 use the available potential of bioenergy, afforestation, and almost all of ERW. Further, biochar is the CDR strategy with the lowest amount of \co\ removal with a cumulative amount of 3-70 \Mco\ from 2025-2050. This aligns with results from our near-optimal solution space of 0-38 \Mco. In Fuhrman et al. \cite{Fuhrman2023}, biochar usage of around 11-25 \Mco\ plays a minor role compared to hundreds of \Mco\ removed by other strategies in Europe.\\ 

Carbon captured from bioenergy in this study stays below 400 \Mco, while Millinger et al. found that up to 1000 \Mco\ are captured from bioenergy for a 5\% system cost increase. In their case, the near-optimal solution space was based on up to 6000 TWh/a biomass potential and 600 \Mco\ underground sequestration potential. The lower capture in our study is partly due to the much lower potentials assumed here of biomass (1386 TWh/a) and of underground sequestration (200 \Mco), and also due to the higher amount of CDR achievable through ERW, perennialization, and afforestation. Consequently, we find that a 5\% system cost increase is enough to reduce carbon capture from bioenergy to 38 \Mco\ compared to more than 250 \Mco\ when DAC is the only replacement option.\\

Existing literature on net-zero emission scenarios includes analyses emphasizing that it is cost-optimal that CDRs are used to offset fossil emissions in difficult-to-abate sectors, and also the opposite, that a fast transformation to Fischer-Tropsch fuels for those sectors is the cost-optimal strategy. Here, by exploring the near-optimal solutions space, we found that a 5\% cost slack allows removal to vary from 262 \Mco\ to 365 \Mco\ and Fischer-Tropsch fuels to represent from 9\% to 76\% of the total oil demand. In essence, the system configuration could be steered towards any of the two strategies within a small cost slack. Taking into account all the uncertainties associated with CDR strategies, their availability should not be taken for granted, and given that they do not represent a clear cost-benefit, future narratives that maximize their usage should be prioritized against those that propose delaying the reduction of fossil emissions with the argument that it is cheaper to offset them. At the same time, most of the carbon management strategies represented here, such as perennialization, afforestation, ERW, underground sequestration, and biogenic carbon capture, are more constrained by their potentials than their costs, and the assumptions about their potentials show wide uncertainties. On top of that, other parameters required to represent them in large-scale systems (cost, energy inputs, scalability limitations) also remain uncertain. Pilot real-world experiences during the next decade could be key to better estimate those parameters and improve their representation in net-zero scenarios.\\

\subsection{Limitations}
Finally, we summarize the main limitations of our study in this paragraph. Our exogenous assumptions regarding the underground sequestration potential, industrial processes and their energy supply, and fuel choices for aviation and shipping shape how some of the carbon management tools can be used. We based our underground sequestration potential on the assumption that it will be needed well into the next century and on planetary boundaries. The planetary limit of \co\ sequestration was recently calculated to be around 42 Gt\co\ for Europe, while the more favoured sedimentary basins with current oil and gas infrastructure account for only around 23 Gt\co\ \cite{Gidden2025}. The conservative use of 200 \Mco\ would deplete these in just over a hundred years, and that does not account for possible hurdles in using some of the sites due to country restrictions or social acceptance. Furthermore, it is possible that the yearly amount will have to be increased after 2050 to reach net negativity in Europe and account for unequal global decarbonization and the possibility that the climate response is asymmetrical and that the same amount of \co\ decreased cools less than its initial warming effect \cite{Zickfeld2021}.\\ 
The industrial sector influences the type and amount of point-source capture used. However, this effect is limited by our findings that none of the individual point-source technologies must necessarily be paired with carbon capture as long as enough carbon is captured in total from point-sources. We assume that oil is used for aviation and methanol for shipping. With the amount of fuel needed and the restricted amount of biomass freely available, most of this demand has to be covered by synthetic fuels to avoid further emissions, leading to a demand for carbon capture and utilization in the system, or has a fossil origin with a corresponding offset. It is possible that endogenizing the provision of heat demand in the industry could free up biomass to produce biofuels. 
The focus of this study was carbon management; therefore, the near-optimal solution space does not capture the full variability of other sectors, e.g. solar and wind capacity or hydrogen use. Variable renewable energy, such as solar and wind, is strongly weather-dependant, and recent research studies the influence of different weather years \cite{Gtske2024}. Here, we only use one weather year and do not assess the effect of interannual weather variability.

\section{Conclusion}
Our analysis shows that Europe can achieve net-zero emissions through multiple combinations of carbon management strategies with only a limited cost increase. All solutions rely, at least in part, on capturing \co\ from bioenergy and remaining fossil combustion or industrial process emissions. Pathways may emphasize either higher use of synthetic fuels or broader deployment of CDR methods. We find that CDR strategies are more constrained by their physical potentials than by their costs, highlighting the need for further pilot studies to better quantify their achievable removal rates and reduce uncertainties.\\

The flexibility observed for a small cost increase should not be interpreted as a justification for delaying decarbonization. Given the physical limits and uncertainties associated with many CDR options, early and sustained emission reductions remain essential to ensure a robust transition. Integrated planning across sectors is therefore critical to balance residual emissions, carbon removal, and fuel needs for a climate-neutral energy system.

\section*{Glossary}
\begin{description}
\item[CC] carbon capture
\item[CCU] Carbon Capture and Utilization
\item[CHP] Combined Heat and Power plant, also known as cogeneration
\item[BECC(S)] Bioenergy with carbon capture (and storage) 
\item[DAC] Direct Air Capture
\item[ERW] enhanced rock weathering 
\item[HSJ] Hop-Skip-Jump
\item[MAA] modelling all alternatives
\item[MGA] modelling to generate alternatives
\end{description}

\section*{CRediT author statement}
Sina Kalweit: Conceptualization (Equal), Methodology (Equal), Software (Lead), Investigation (Lead),
Writing -- original draft (Lead), Writing -- Review \& Editing (Equal), Visualization (Lead)\\
Alberto Alamia: Methodology (Supporting), Software (Supporting), Writing -- Review \& Editing (Equal)\\
Ricardo Fernandes: Methodology (Supporting), Software (Supporting), Writing -- Review \& Editing (Supporting)\\
Marta Victoria: Conceptualization (Equal), Methodology (Equal), Software (Supporting), Writing -- Review
\& Editing (Equal), Supervision (Lead)

\section*{Acknowledgments}
S.K. and R.F. are fully funded by The Novo Nordisk Foundation CO$_2$ Research Center (CORC) under grant number CORC005. A.A. is fully funded by Villum Fonden under grant number 40519. 

\section*{Data and code availability}
The model is implemented with the open energy modeling framework PyPSA v0.31.0 and based on the model PyPSA-Eur v2025.01.0, which was adapted to this paper as can be seen in \url{https://github.com/ricnogfer/pypsa-eur/tree/v2025.01.0_co2_stores}. It uses the costs and technology assumptions included in the technology-data v0.10.1, which was adapted to this paper as can be seen in \url{https://github.com/BertoGBG/technology-data/tree/CO2_stores_latest}.\\
The code for the MGA approach, datasets and visualization scripts can be accessed from the public repository: 
10.5281/zenodo.19070065


\bibliographystyle{elsarticle-num} 
\bibliography{bib_merged}

\clearpage
\FloatBarrier
\include{Supplementary_mga}

\end{document}

%% file: Supplementary_mga.tex
\setcounter{figure}{0}
\renewcommand{\figurename}{Fig.}
\renewcommand{\thefigure}{S\arabic{figure}}
\setcounter{table}{0}
\renewcommand{\thetable}{S\arabic{table}}
\setcounter{section}{0}
\renewcommand{\thesection}{Supplementary Note \arabic{section}}
\setcounter{equation}{0}
\renewcommand{\theequation}{S\arabic{equation}}
\noindent\textbf{\large{Supplementary Information}}\\
\label{app1}
\section{Supplementary}
\subsection{Carbon management}
\label{Sup:carbonmanagement}
The \co\ that is captured from point-source emissions or conversion processes $C_{cap}$ and is available for underground sequestration or conversion to synthetic fuels is not equal to how the carbon capture balances \co\ emissions in the atmosphere $C_{bal}$. 
\begin{align}
  C_{cap} &= \sum_n(p\cdot c_{r,n})\\
  C_{bal,biomass} &= -\left(\sum_{n_{combustion}}(c_{s,n}\cdot p)+\sum_{n_{conversion}}(c_{s,n}-(1-p)c_{r,n})\right)\\
  C_{bal,fossil} &= \sum_{n}((1-p)c_{s,n})\\
  C_{bal,biogas} &= -\sum_{n}(c_{s,n}+p\cdot c_{r,n})
\end{align}
where $p$ is the percentage captured, $c_{r,n}$ the amount of \co\ released during the process, and $c_{s,n}$ the amount of \co\ stored in the product per node $n$. For instance, when biogas is converted to methane, $c_{s,n}=0.198tCO_2/MWh_{th}$, $p=90\%$ and $c_{r,n}=0.0868tCO_2/MWh_{th}$, which means that for every $0.28tCO_2/MWh_{th}$ abated, only $0.08tCO_2/MWh_{th}$ are captured as a usable \co\ source.
\subsection{MGA method}
For the integer set, the first vector $v_1$ is sampled from [-1,1] and then inverted to ensure that every variable is maximized and minimized at least once. The draw from the integer set [-1,0,1] is checked to ensure it is unique and has not been solved before. In cases where drawing 100 times from the integer set does not lead to a new unique set, the vector $v$ is drawn from the uniform distribution. In addition to this method, we also manually included scenarios where every relevant technology is minimized. Here, we use the capacities of the carbon capture, conversion, and sequestration technologies as well as the CDR strategies from Tab. \ref{tab:CDR} as decision variables. The capacities of all nodes of a technology get the same weight of $v$ assigned to explore the variety of the total use of a technology.
\label{sup:vol}
\begin{figure}[h]
\centering
\includegraphics[width=0.6\textwidth]{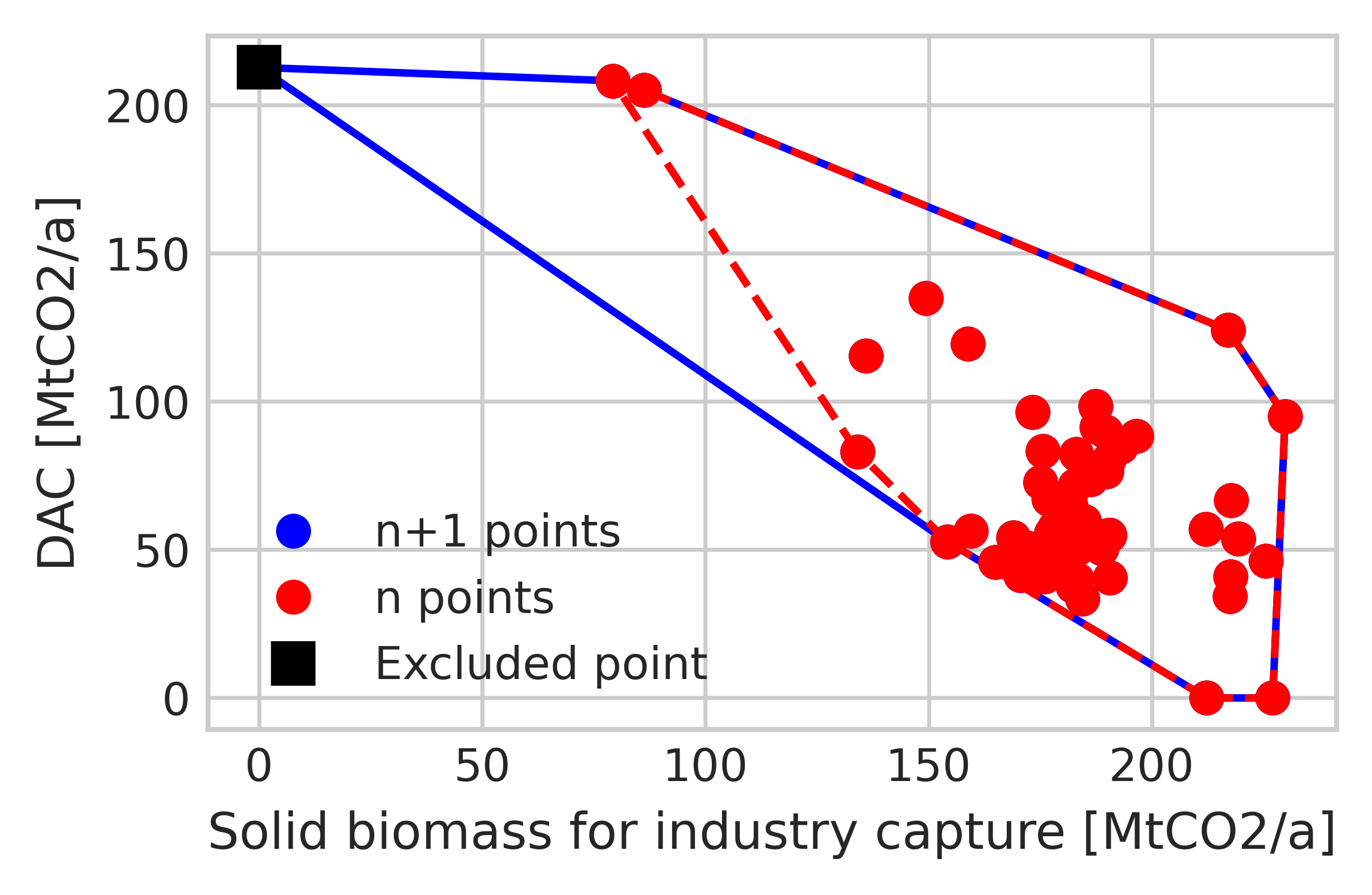}
\caption{Example for volume expansion in two dimensions}
\label{Fig:vol}
\end{figure}
An evaluation criterion of the MGA method is finding enough diverse solutions to ensure that the variety of possible solutions is sufficiently mapped. A way to measure this is by calculating the change of volume spanned when adding new solutions. It is considered that enough solutions are found, when the change of volume when adding a new solution is negligible. Following \cite{Lau2024}, summing the projection to the 2D-subspaces of each unique pairwise combination of the selected variables gives a computationally feasible approximation of the expansion of the near-optimal solution space. An example of the 2D-subspaces is shown in Fig. \ref{Fig:vol}, where the volume expands from $n$ to $n+1$ points, because solution $n+1$ increases the capacity of perennialization. The volume is calculated after every five new results and as a convergence criterion, the difference of the most recent volume to the last three calculated volumes has to be lower than 1\%.

\subsection{Overview CDR potentials and costs}
\begin{table}[!htb]
    \renewcommand{\arraystretch}{0.8}
    \small
    \centering
    \caption{Land availability and CO$_2$ removal/sequestration potential}
    \label{table_cdrs_land_and_sequestration_potentials}
    \begin{tabular}{p{3cm}cc}
        \toprule
        \textbf{CDR}              & \textbf{Land availability [Mha]} & \textbf{Rem./seq. potential [MtCO$_2$/a]} \\
        \midrule
        Underground seq. & -                                & 200                                       \\
        Afforestation             & 15                               & 83                                        \\
        Perennialization                & 21                               & 42                                        \\
        Biochar                   & 23                               & 30                                        \\
        ERW                        & 14                               & 42                                        \\
        \bottomrule
    \end{tabular}
\end{table}

\vspace{2cm}

\begin{table}[!htb]
    \renewcommand{\arraystretch}{0.7}
    \setlength{\tabcolsep}{3pt} 
    \smallskip
    \footnotesize 
    \centering
    \caption{Main parameters used to model the different CDRs.}
    \label{tab:cdrs_parameters}

    \begin{tabular}{l *5{S} l}
        \toprule
        \textbf{Parameter}  &
        \textbf{Geol. seq.} &
        \textbf{Aff.} &
        \textbf{Per.} &
        \textbf{Biochar} &
        \textbf{ERW} &
        \textbf{Unit}  \\
        \midrule

        Electricity in          &
        \multicolumn{1}{c}{--} & \multicolumn{1}{c}{--} & 0.32 & 0.32 & 0.19 & [MWh/tCO$_2$]\\

        Biomass in  &
        \multicolumn{1}{c}{--} & \multicolumn{1}{c}{--} & \multicolumn{1}{c}{--} & 7.67 & \multicolumn{1}{c}{--} &
        [MWh$_{biomass}$/tCO$_2$]\\

        Basalt in     &
        \multicolumn{1}{c}{--} & \multicolumn{1}{c}{--} & \multicolumn{1}{c}{--} & \multicolumn{1}{c}{--} & 3.33 &
        [t$_{basalt}$/tCO$_2$] \\

        Biogas out            &
        \multicolumn{1}{c}{--} & \multicolumn{1}{c}{--} & 0.86 & \multicolumn{1}{c}{--} & \multicolumn{1}{c}{--} &
        [MWh/tCO$_2$]  \\

        Heat out                &
        \multicolumn{1}{c}{--} & \multicolumn{1}{c}{--} & \multicolumn{1}{c}{--} & 3.79 & \multicolumn{1}{c}{--} 
        &[MWh/tCO$_2$]\\

        Capital cost             &
        \multicolumn{1}{c}{--} & \multicolumn{1}{c}{--} & 6.05e6 & 8.94e6 & \multicolumn{1}{c}{--} &  [EUR/(tCO$_2$/h)] \\

        FOM                    &
        \multicolumn{1}{c}{--} & \multicolumn{1}{c}{--} & \multicolumn{1}{c}{--} & 3.42 & \multicolumn{1}{c}{--} &
        [\% capital cost]\\

        VOM                          &
         \multicolumn{1}{c}{--} &  \multicolumn{1}{c}{--} & 190 & 48 & 190 & [EUR/tCO$_2$]\\
        avg. seq. cost                          &
        10 & 113 & \multicolumn{1}{c}{--} & \multicolumn{1}{c}{--} & \multicolumn{1}{c}{--} & [EUR/(tCO$_2$/h)]\\
        \bottomrule
    \end{tabular}
    \caption*{\footnotesize
\textit{Notes:} Geol. seq. = Underground sequestration; Aff. = Afforestation; Per. = Perennialization; Bio. = Biochar; ERW = Enhanced rock weathering.}
\end{table}

\section{Additional solutions from near-optimal solution space}
\subsection{DAC can only partly replace point-source carbon capture}
\begin{figure}
  \includegraphics[width=\textwidth]{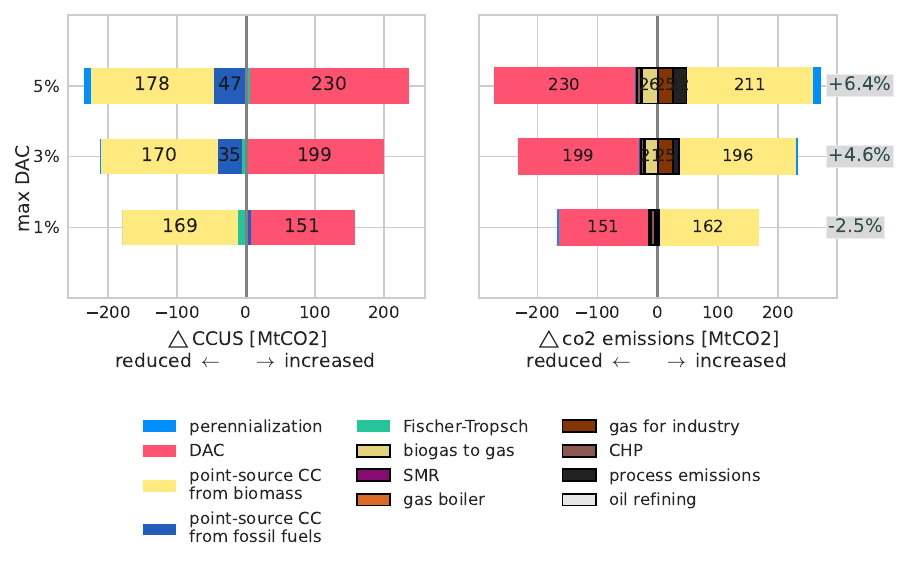}
\caption[Solution space: maximizing DAC]{Differences in technology use for scenario where DAC is maximized, left: balance of decision variables, right: \co\ emission balance. Percentages show how the total amount of emissions changes compared to the optimal solution.}
\label{Fig:ex3}
\end{figure}
For a 1\% budget increase with maximized DAC use, CDR strategies, \co\ conversion and sequestration remain unchanged compared to the optimal solution. DAC use can increase up to 213 \Mco\ for the 1\% budget increase, replacing 169 \Mco\ captured by biomass point-source, see Fig. \ref{Fig:ex3}. This is 75\% of the total amount of point-source capture from biomass in the optimal solution, but even for the 5\% budget increase, DAC cannot fully replace point-source capture. Since both DAC and point-source carbon capture from biomass combined with underground sequestration result in net \co\ removals, it is expected that they are substitutable to a certain degree. For higher budget, additional point-source capture from fossil fuel is replaced by DAC which leads to an increase in emissions from point-source without carbon capture from exogenously set gas for industry demand and process emissions. These are partially offset by increasing biogas to gas conversion although total amount of emissions also increases. 
\begin{figure}[htbp!]
             \centering
             \includegraphics[clip, trim=0cm 0cm 0cm 0cm,width=\textwidth]{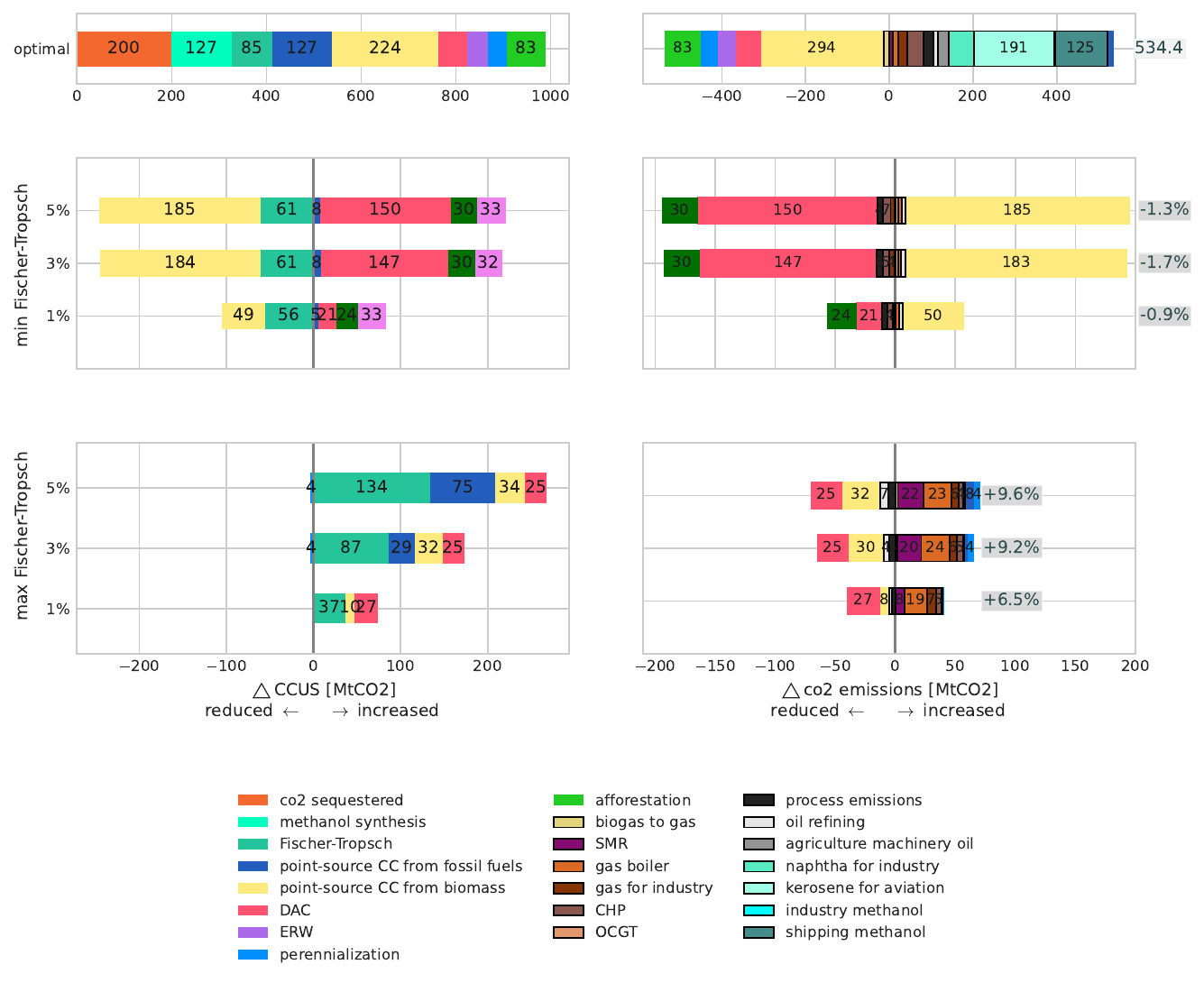}
\caption{CCUS for Fischer-Tropsch minimization and maximization}
\label{fig:exFT}
\end{figure}

\begin{figure}[htbp!]
             \centering
             \includegraphics[clip, trim=0cm 0cm 0cm 0cm,width=\textwidth]{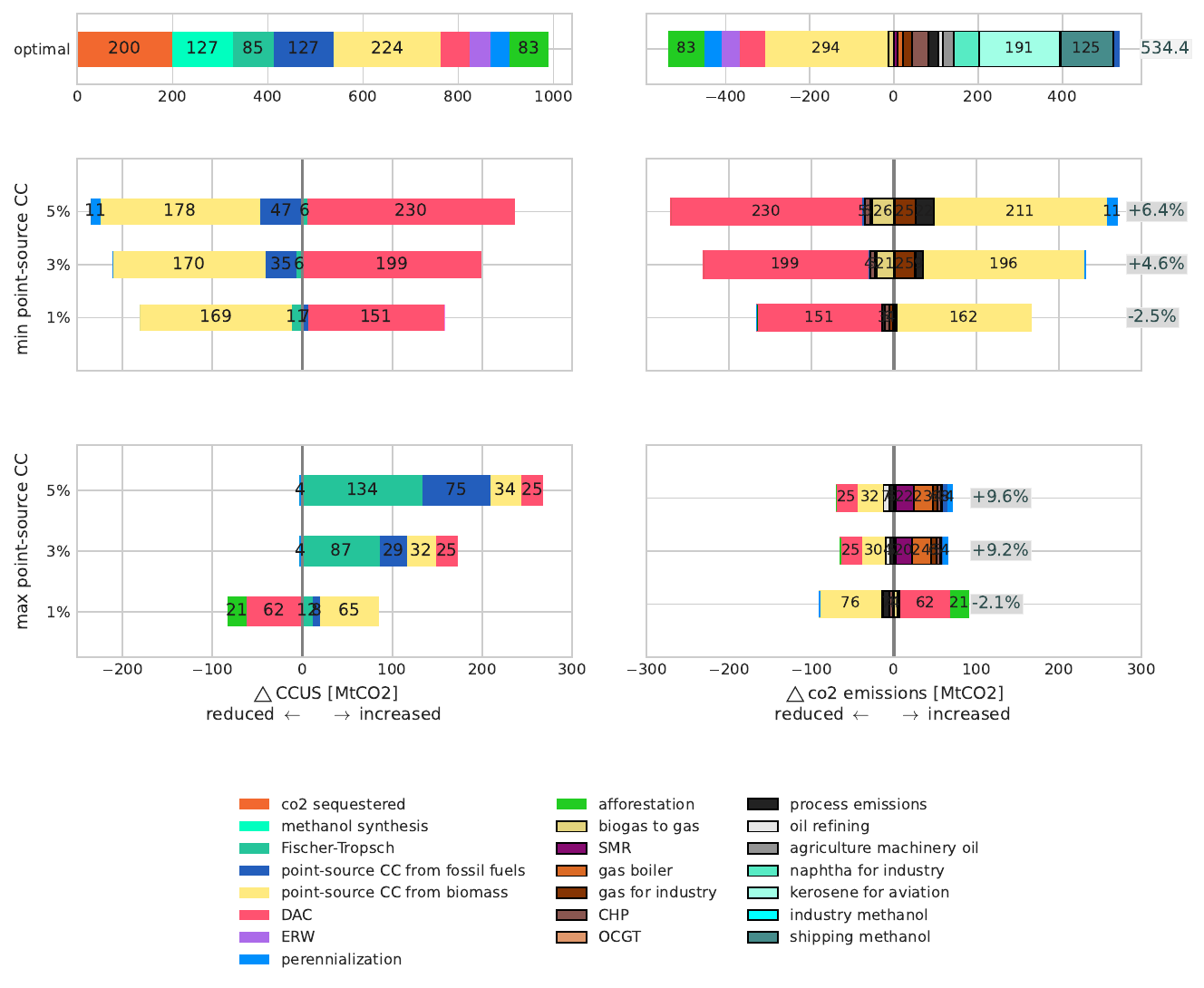}
\caption{CCUS for Point-source carbon capture minimization and maximization}
\label{fig:exPS}
\end{figure}

 \begin{figure}[htbp!]
             \centering      
             \includegraphics[clip, trim=0cm 0cm 0cm 0cm,width=\textwidth]{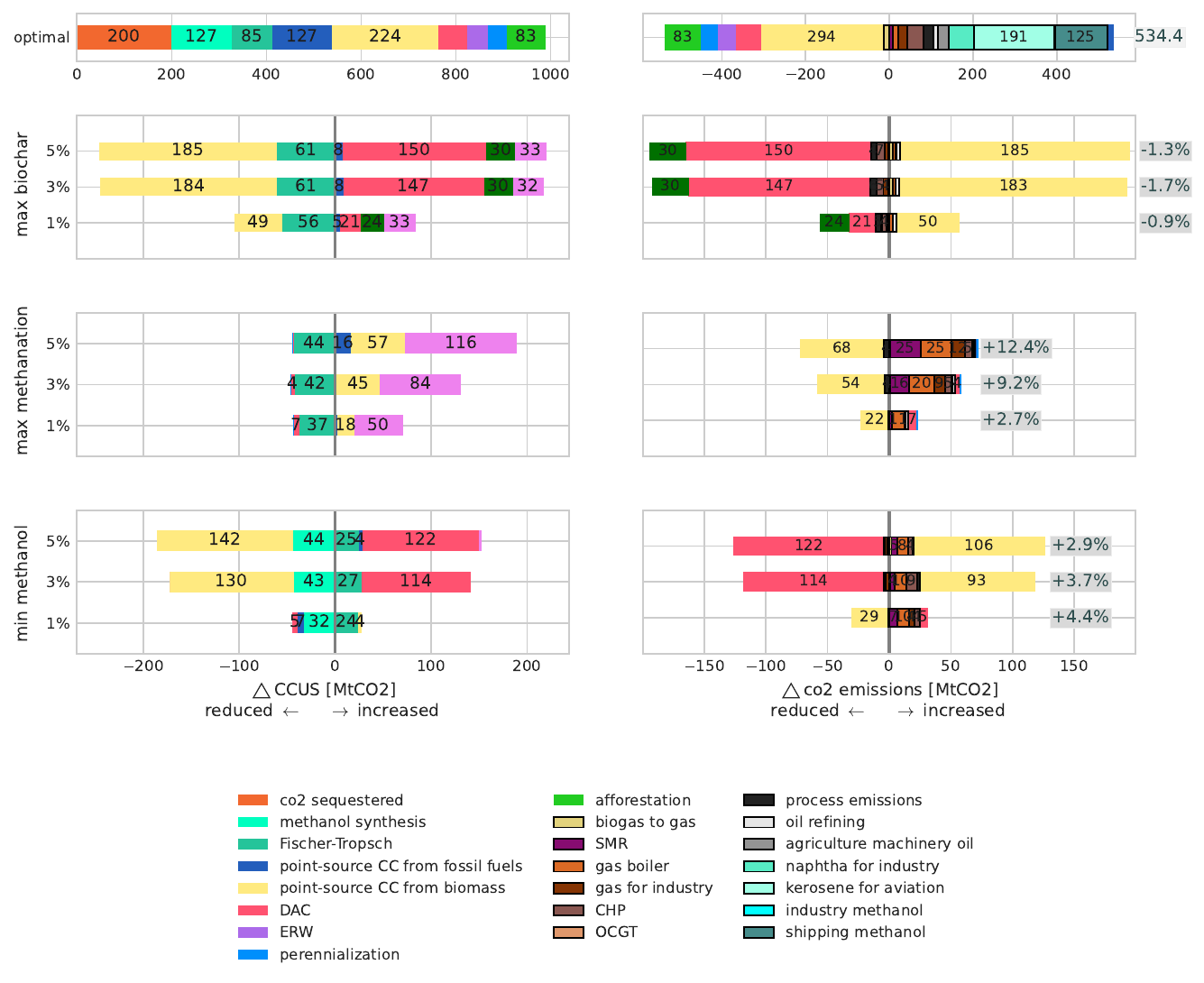}
\caption{CCUS for biochar maximization, methanation maximization, methanol synthesis minimization}
\label{fig:exbmm}
\end{figure}
 
\begin{figure}[htbp!]
\centering
            \includegraphics[width=0.9\textwidth]{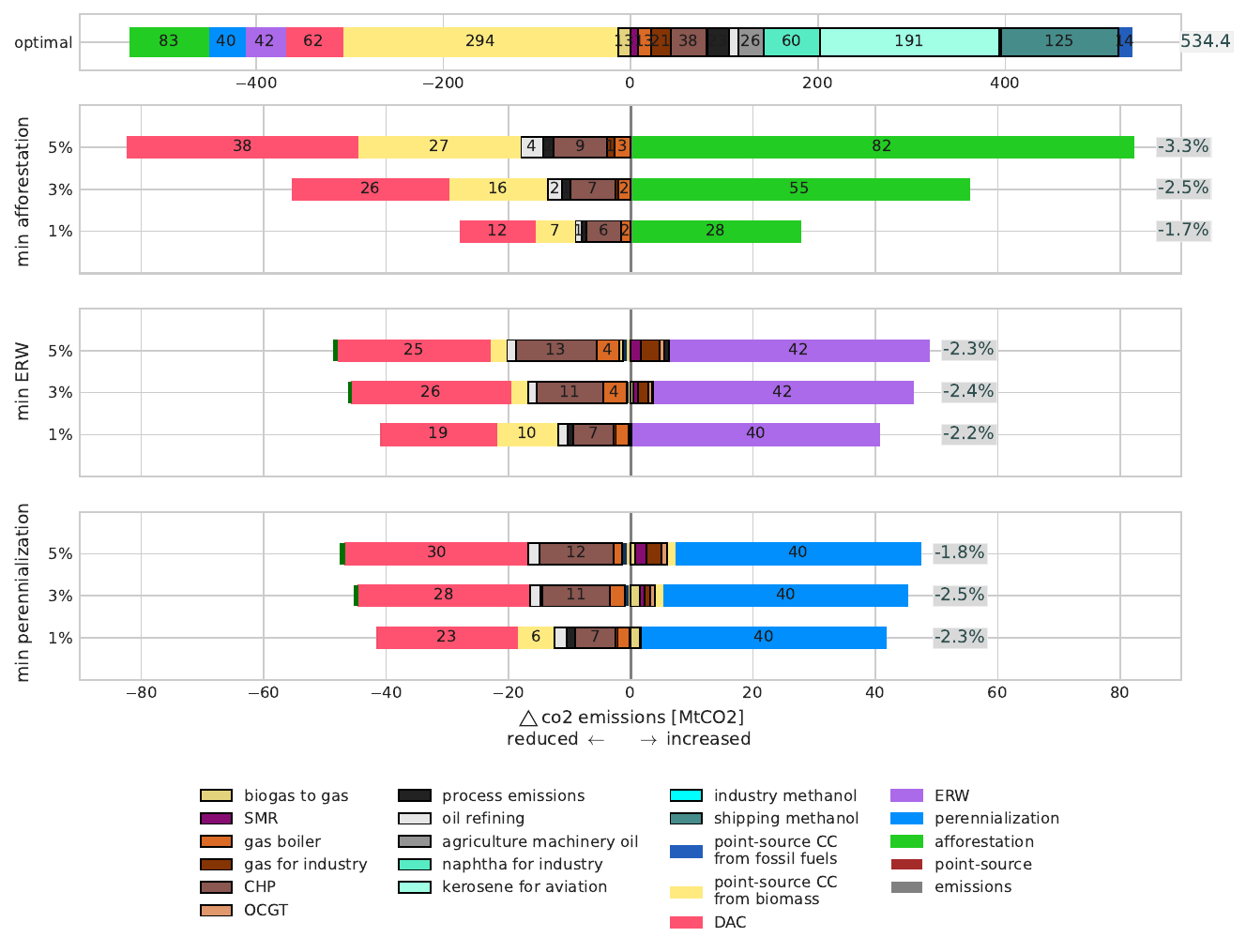}
\caption{\co\ emission balance when reducing afforestation, ERW or perennialization.}
\label{fig:exewper}
\end{figure}
\FloatBarrier
\subsection{Temporal resolution}
\begin{figure}[htbp!]
\centering
         \begin{subfigure}[b]{0.45\textwidth}
          \centering
            \includegraphics[width=\textwidth]{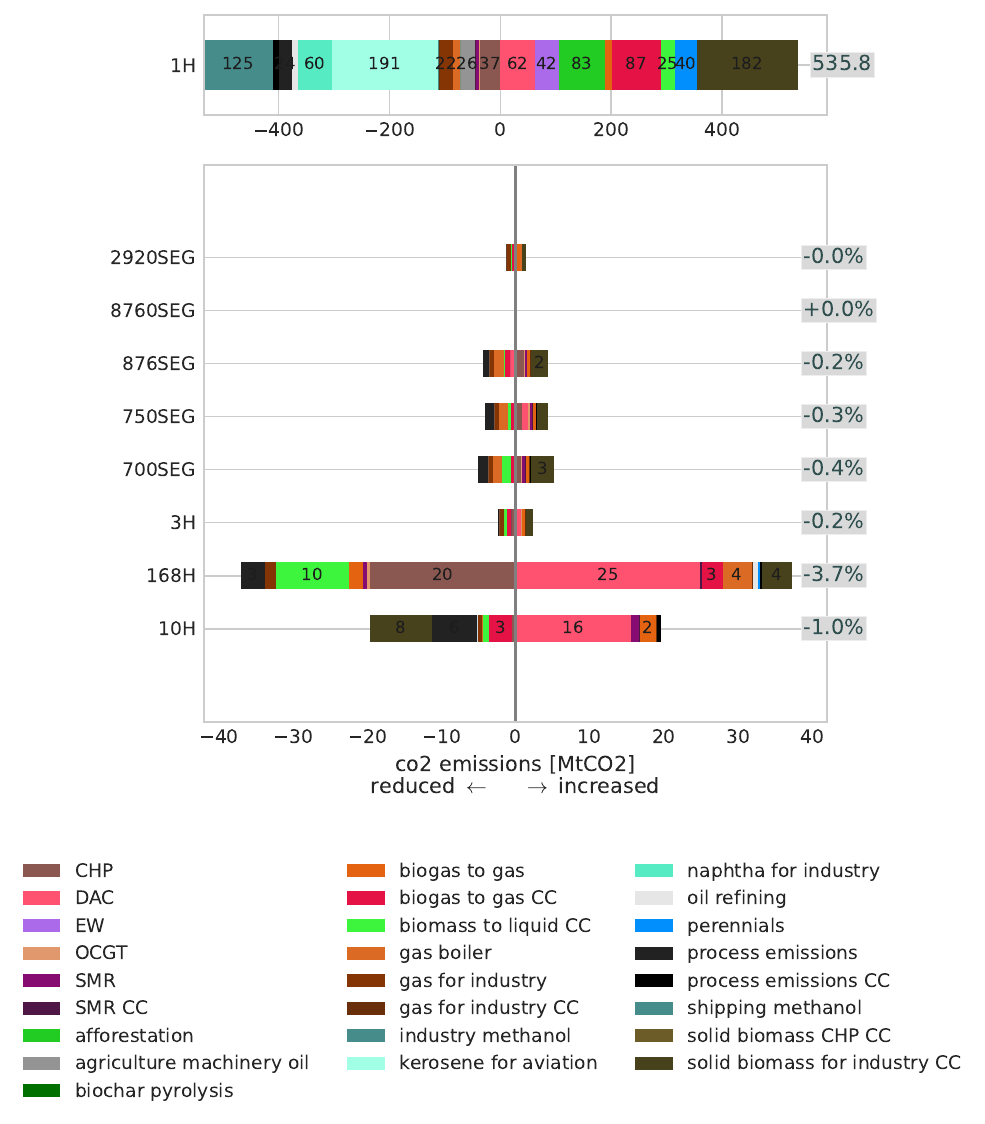}
         \end{subfigure}
        \begin{subfigure}[b]{0.45\textwidth}
         \centering
          \includegraphics[width=\textwidth]{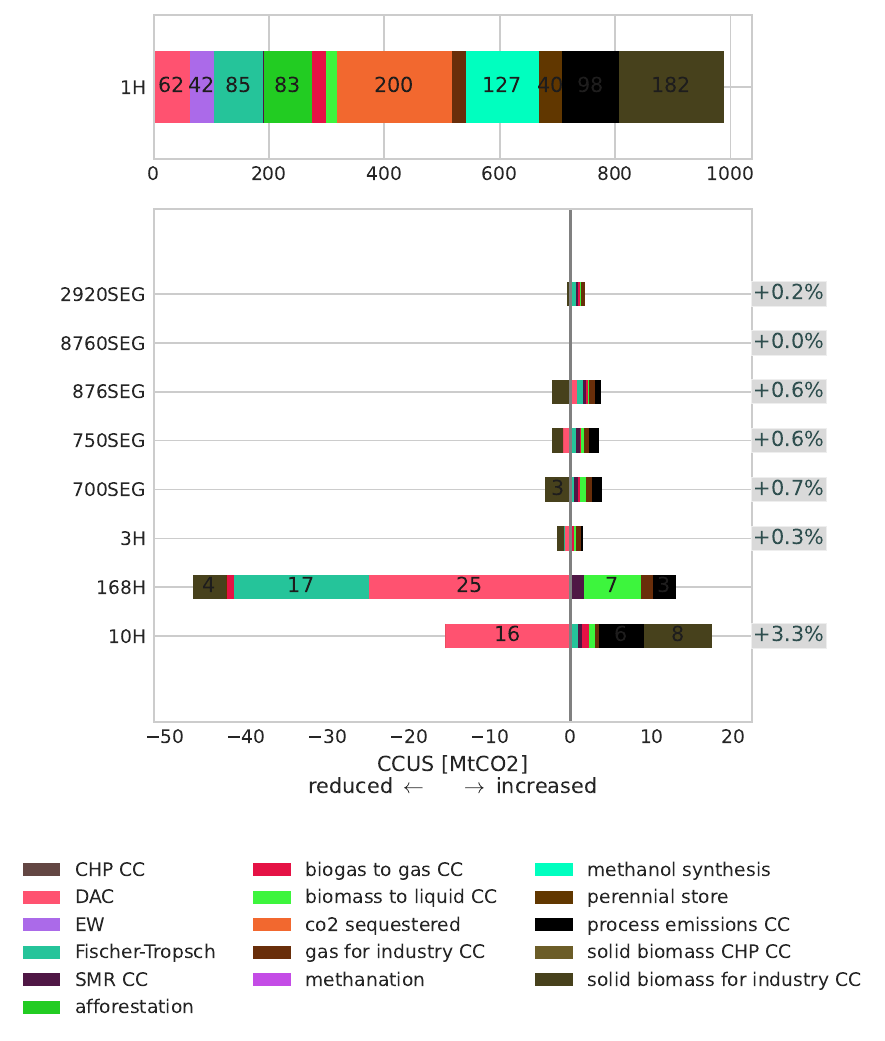}
         \end{subfigure}
\caption{Comparison of carbon management for different temporal resolutions.}
\label{fig:tempcomp}
\end{figure}
\FloatBarrier
\subsection{Spatial distribution}
 \begin{figure}[htbp!]
\centering

            \begin{subfigure}[b]{\textwidth}
             \centering
             \includegraphics[clip, trim=0cm 0cm 0cm 0cm,width=\textwidth]{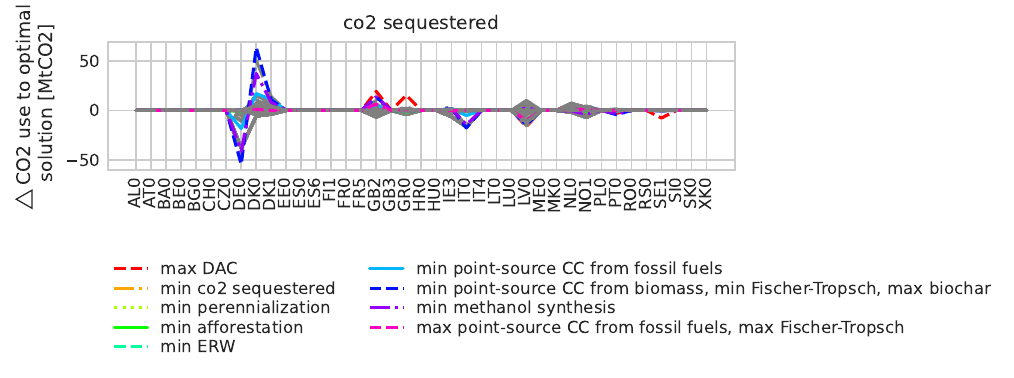}
             \label{fig:spac}
         \end{subfigure}
\caption{Differences in CCUS per node compared to cost-optimal solution.}
\label{fig:spatial3}
\end{figure}

 \begin{figure}[htbp!]
\centering
    \begin{subfigure}[b]{\textwidth}
             \centering      
             \includegraphics[clip, trim=0cm 0cm 0cm 0cm,width=\textwidth]{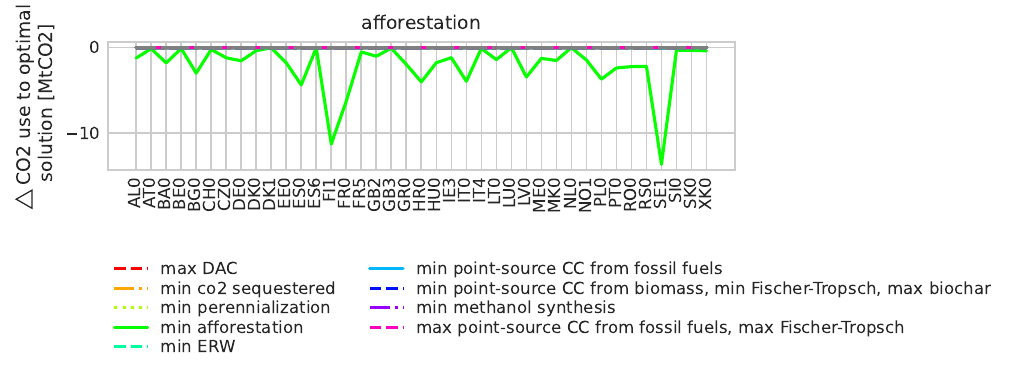}
             \label{fig:spa}
         \end{subfigure}
            \begin{subfigure}[b]{\textwidth}
             \centering
             \includegraphics[clip, trim=0cm 0cm 0cm 0cm,width=\textwidth]{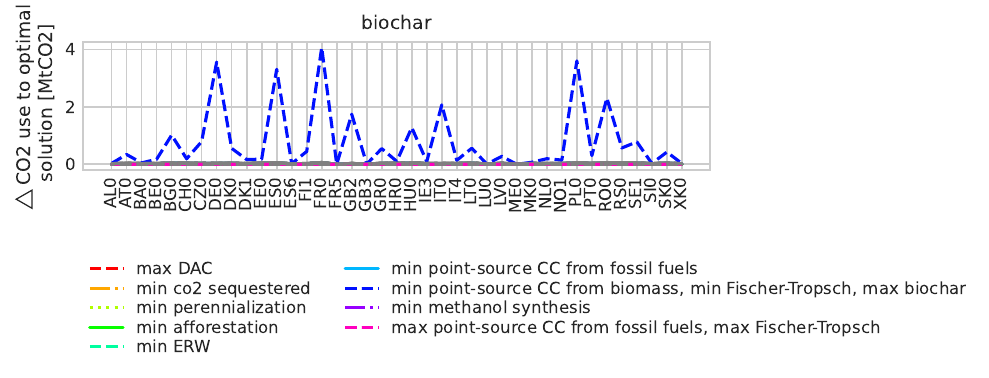}
             \label{fig:spb}
         \end{subfigure}
         \begin{subfigure}[b]{\textwidth}
             \centering
             \includegraphics[clip, trim=0cm 0cm 0cm 0cm,width=\textwidth]{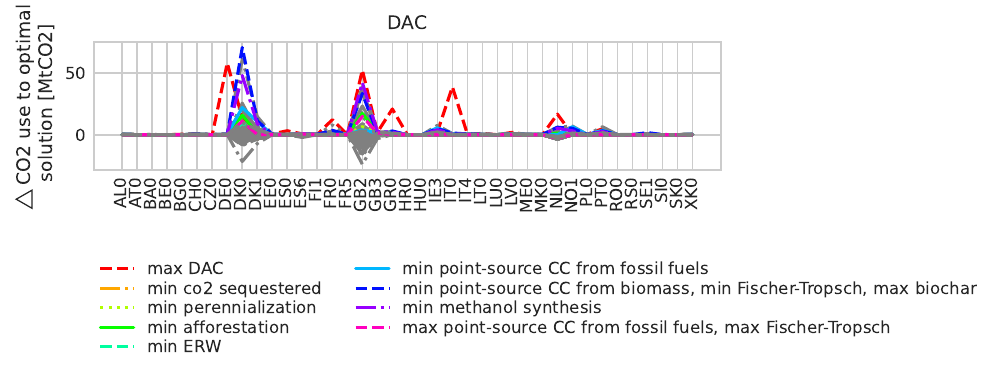}
             \label{fig:spD}
             \end{subfigure}
\caption{Spatial distribution for afforestation, biochar, DAC}
\label{fig:spatial0}
\end{figure}

 \begin{figure}[htbp!]
\centering
            \begin{subfigure}[b]{\textwidth}
             \centering
             \includegraphics[clip, trim=0cm 0cm 0cm 0cm,width=\textwidth]{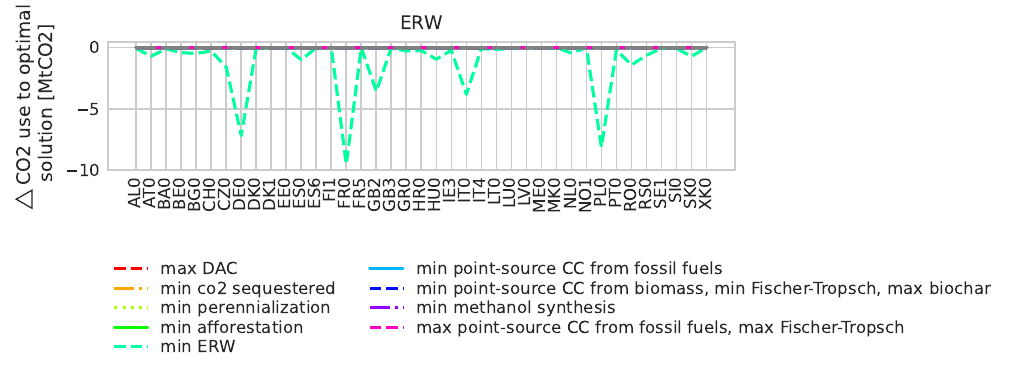}
             \label{fig:spaE}
         \end{subfigure}
         \begin{subfigure}[b]{\textwidth}
             \centering      
             \includegraphics[clip, trim=0cm 0cm 0cm 0cm,width=\textwidth]{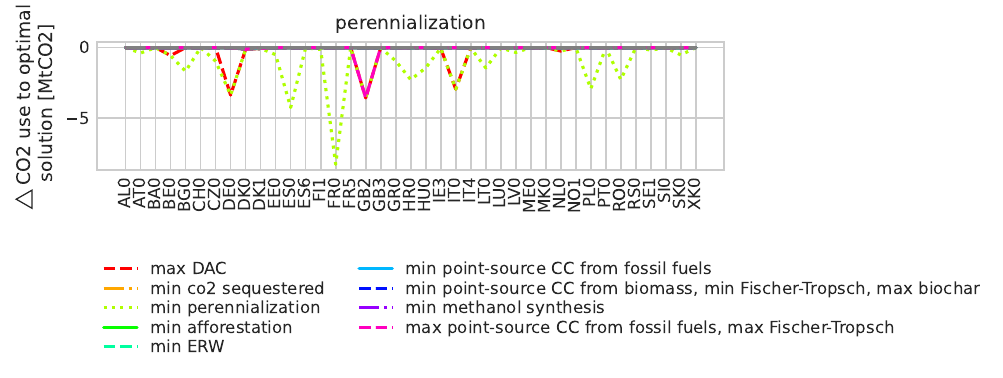}
             \label{fig:spap}
         \end{subfigure}
\centering
        \begin{subfigure}[b]{\textwidth}
             \centering      
             \includegraphics[clip, trim=0cm 0cm 0cm 0cm,width=\textwidth]{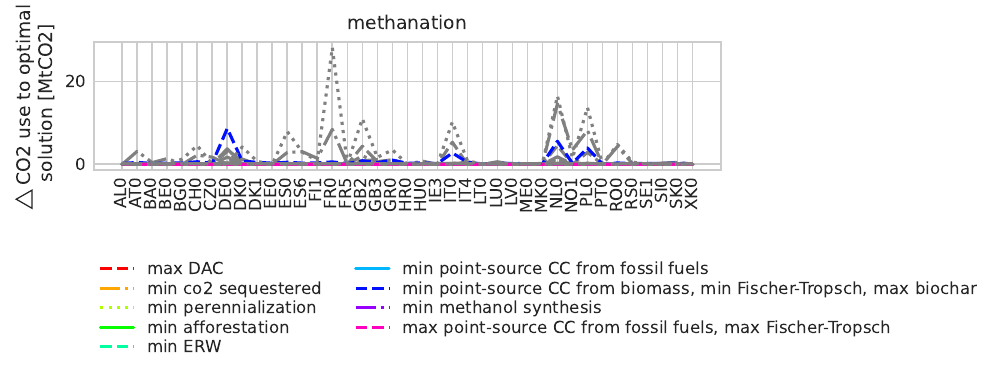}
             \label{fig:spam}
         \end{subfigure}
    \caption{Spatial distribution for ERW, perennialization, methanation}
\label{fig:spatial1}
\end{figure}

 \begin{figure}[htbp!]
\centering
         \begin{subfigure}[b]{\textwidth}
             \centering      
             \includegraphics[clip, trim=0cm 0cm 0cm 0cm,width=\textwidth]{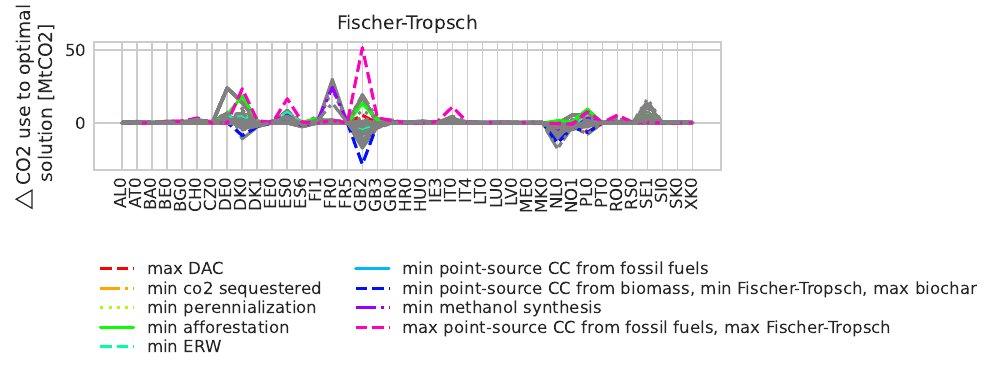}
             \label{fig:spaT}
         \end{subfigure}
         \begin{subfigure}[b]{\textwidth}
             \centering      
             \includegraphics[clip, trim=0cm 0cm 0cm 0cm,width=\textwidth]{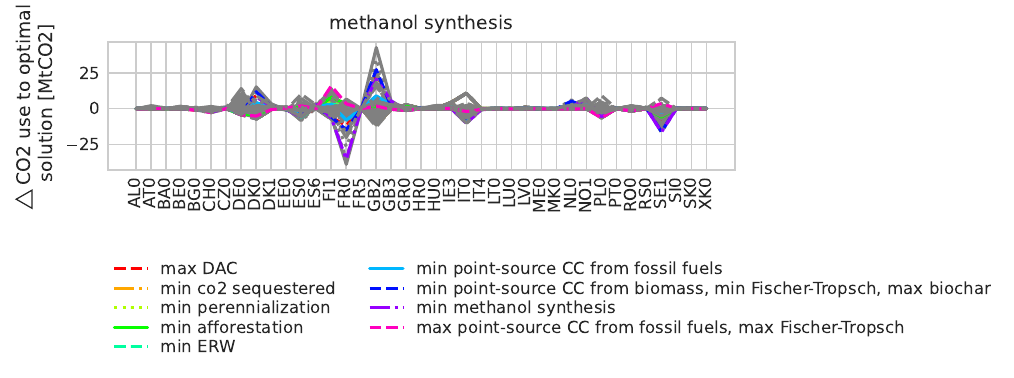}
             \label{fig:spame}
         \end{subfigure}
         \begin{subfigure}[b]{\textwidth}
             \centering
             \includegraphics[clip, trim=0cm 0cm 0cm 0cm,width=\textwidth]{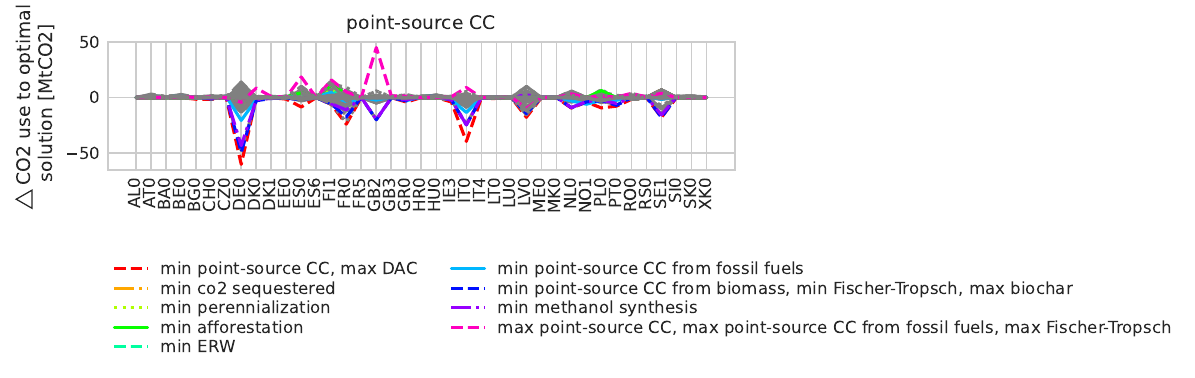}
             \label{fig:spapo}
             \end{subfigure}
\caption{Spatial distribution for Fischer-Tropsch, methanol synthesis, point-source CC}
\label{fig:spatial2}
\end{figure}

\subsection{Producing syngas from solid biomass (BioSNG)}
The conversion from biomass to BioSNG is more efficient leading to 112 TWh biomass converted to 70TWh BioSNG CC when allowing BioSNG and BioSNG CC compared to 104 TWh biomass converted to only 39 TWh biomass to liquid CC in the optimal solution of this paper. The system then uses the higher amount of biogas that replaced natural gas to reduce gas for industry CC and Fischer-Tropsch and increase the use of fossil oil.
This leads to a total system cost that is 0.05\% cheaper when BioSNG is activated. With this small difference the cost-optimal solution is highly susceptible to any kind of changes in the assumptions, especially when loosening restrictions around biomass availability and \co\ underground sequestration. Both reduce the importance of the efficient conversion of biomass to biofuels of any kind to reduce emissions.
For example, when increasing the \co\ underground sequestration to 600 MtCO2/a (which is the lowest potential in Millinger et al.) the cost-optimal solution only uses biomass to liquid CC.